\documentclass[twoside]{article}
\usepackage{fleqn,espcrc2}

\usepackage{graphicx}
\usepackage[figuresright]{rotating}

\newcommand{\AmS}{{\protect\the\textfont2
  A\kern-.1667em\lower.5ex\hbox{M}\kern-.125emS}}

\title{Overlap Fermions on a $20^4$ Lattice}

\author{
K. F. Liu
\address{Department of Physics and Astronomy, 
University of Kentucky, Lexington, KY 40506, USA} 
\thanks{Talk presented by K.F. Liu at Lattice 2000, Bangalore, India},
S. J. Dong$^{\rm \;a}{}$, 
F. X. Lee
\address{Department of Physics, George Washington University, 
Washington, DC 20052, USA}{}
\address{Jefferson Lab, 12000 Jefferson Avenue, Newport News, VA 23606, USA},
and J. B. Zhang$^{\rm \;a}$   
\address{Zhejiang Institute of Modern Physics, Zhejiang University, 
Hangzhou 310027, P.R. China} }

\begin{document}

\begin{abstract}
We report results on hadron masses, fitting of the quenched chiral log, and
quark masses from Neuberger's overlap fermion on a quenched $20^4$ lattice with
lattice spacing $a = 0.15$ fm. We used the improved gauge action which is shown
to lower the density of small eigenvalues for $H^2$ as compared to the Wilson
gauge action. This makes the calculation feasible on 64 nodes of  CRAY-T3E.
Also presented is the pion mass on a small volume ($6^3 \times 12$
with a Wilson gauge action at $\beta = 5.7$). We find that for configurations
that the topological charge $Q \ne 0$, the pion mass tends to a constant and
for configurations with trivial topology, it approaches zero possibly linearly
with the quark mass. 
\vspace{1pc}
\end{abstract}

\maketitle

   Overlap fermion has great promise in studying chiral symmetry on the
lattice~\cite{neu00}. Recent numerical implementation of the overlap fermions 
has led to studies of chiral condensate~\cite{ehn99b,hjl99}, quark
mass~\cite{dll00}, as well as the checking of chiral symmetry and
scaling~\cite{dll00,eh00}. However these studies are limited to 
small volumes due to the enormous numerical cost associated with approximating
the matrix sign function which is about fifty times more than
inverting the quark matrix of the vanilla Wilson action. 

   In the optimal rational approximation~\cite{ehn99a} of the matrix sign
function of Neuberger's overlap operator~\cite{neu98}
\begin{equation}  \label{neu}
D(m_0)=  1 + \frac{m_0a}{2} + (1 - \frac{m_0a}{2} ) \gamma_5 \epsilon (H),
\end{equation}
where $\epsilon (H) = H /\sqrt{H^2}$ and $H$ is taken to be the hermitian
Wilson-Dirac operator, i.e. $H = \gamma_5 D_w$ with $0.25 > \kappa > \kappa_c)$,
it is cost effective to project out a relatively few eigenmodes with very small
eigenvalues in the operator $H^2$ in order to reduce the condition number and
speed up the convergence in the inner do loop~\cite{ehn99b,dll00}. At the same
time, it improves the chiral symmetry relation such as the
Gell-Mann-Oakes-Renners relation~\cite{dll00}. However, it is
shown~\cite{ehn99c} that the density of these small eigenmodes grows 
as $e^{\sqrt{a}}$ with $a$ being the lattice spacing.  As a result, it is very
costly and impractical to work on large volumes with the currently used lattice
spacings. There are simply too many small eigenmodes to be projected out. 

   For this reason, we explore other options to clear this
hurdle.  We have tested the $O(a_s^2)$ improved gauge action of
Morningstar and Peardon~\cite{mp97} and find that the density of these small
eigenvalue modes is decreased to a point that it becomes feasible to go to
large volumes with a size of the lattice 3 to 4 times of the Compton
wavelength of the lightest pion. We further find that the anisotropic
action~\cite{mp97} requires projection of more small eigenvalues in $H^2$ in
order to achieve the same convergence in the inner loop than does the
isotropic one. Thus, we use the isotropic action. We also find that
using the clover action with either sign requires the projection of more small
eigenvalue modes. Therefore we use the Wilson action for $H$ in the 
Neuberger operator with $\kappa = 0.19$. On a $20^4$ lattice with $a = 0.15$ fm
as determined from the Sommer scale $r_0$, we project out 80 small eigenvalues. 
Beyond these eigenmodes, the level density becomes dense. As a result, the
number of conjugate gradient steps for the inner loop is about 160 and the
conjugate gradient steps is about 210 for the outer loop. These numbers are
about the same as those for the Wilson gauge action on small
volumes~\cite{dll00}. Therefore, other than the overhead of projecting out the
small eigenvalue modes, the cost scales  linearly with volume.

   We shall report preliminary results on 38 gauge configurations. Given $a =
0.15$ fm for the $20^4$ lattice, the physical length of the lattice is 3 fm. We
shall study pion mass as low as $\sim 240$ MeV so that the size of the lattice
is about 3.6 times of the pion Compton wavelength. 

\begin{figure}[tbh]
\vspace*{5.0cm}
\includegraphics{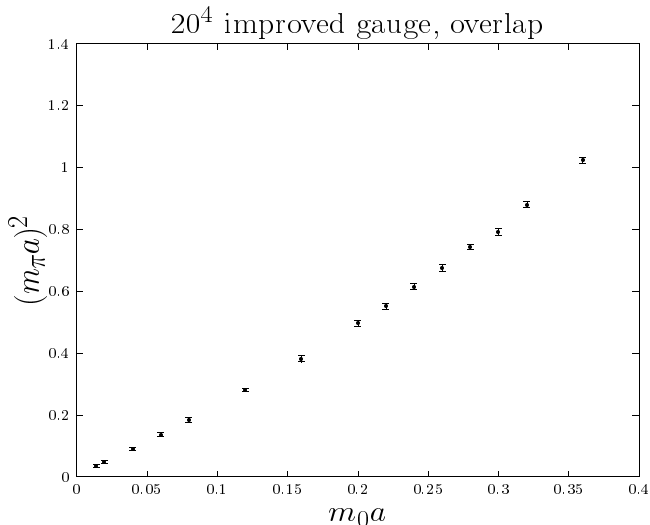} \caption{Pion mass squared as a function of the bare quark mass
$m_0a$ on the $20^4$ lattice with $a = 0.15$ fm.} \label{pimass2}
\end{figure}

   We first show the pseudoscalar mass squared $m_P^2 a^2$ as a function of
$m_0 a$ in Figure~\ref{pimass2}. We fit them in the following form~\cite{sbg92} with and
without the quenched chiral log term $\delta$
\begin{equation}  \label{chi_log}
m_P^2 a^2 = c + 2A m_0 a^2\{1 -\delta \ln(2Am_0/\Lambda_{\chi}^2)\} + 4B m_0^2.
\end{equation}

As shown in Table~\ref{table:1}, the fits with and without
the quenched chiral log $\delta$ give comparable $\chi^2$/DF for the 
6 to 10 smallest quark masses ranging from 0.014 to 0.24 (see Figure~\ref{pimass2}).
 The fit is insensitive to $\Lambda_{\chi}$ in the range of 0.6 to 1.4
GeV. We list results with $\Lambda_{\chi} = 1.0 $ GeV in Table~\ref{table:1}.  
Furthermore, the errors on  $\delta$ are much larger than the corresponding
central values. We conclude that there is no signal for the quenched chiral log
in the range $m_{\pi}/m_{\rho} \sim 0.35 - 0.80$. This is in contrast to the
study of the Wilson fermion action on lattices with the same physical volume and
comparable quark masses~\cite{abb00} which supports the presence of quenched
chiral logarithms with a magnitude $\delta \sim 0.1$. We also note that in our
fit, the intercept c is consistent with zero which is to be expected.

\begin{table*}[ht]
\caption{The fitted parameters in Eq. (\ref{chi_log}). $\delta = 0$ 
corresponds to the case without the chiral log. } \label{table:1}
\newcommand{\m}{\hphantom{$-$}}
\newcommand{\cc}[1]{\multicolumn{1}{c}{#1}}
\renewcommand{\tabcolsep}{2pc} 
\renewcommand{\arraystretch}{1.2} 
\begin{center}
\begin{small}
\begin{tabular}{@{}llllll}
\hline
\# pts. & c  & $\delta$ & A  & B & $\chi^2$/DF  \\ 
\hline   
   5 & 0.0016(50) & 0  & 4.51(22) &  0 & 0.008  \\
   6 & -0.0006(41) & 0 & 4.65(13) &  0 & 0.2  \\
   6 & 0.0045(70) & 0  & 4.17(56) & 12(24) & 0.02 \\
   6 &            &       &        &             \\
   8 & 0.0047(53) & 0   & 4.15(30)  & 7.5(29) &  0.03 \\
   8 & 0.004(18)  & 0.015(297) & 4.13(56) & 8.2(149) & 0.035 \\
  10 & 0.0051(49) & 0  & 4.12(22)  & 7.8(20)  & 0.04  \\
  10 & 0.02(15)   & 0.043(227) & 4.07(36) & 9.7(98) & 0.04  \\
\hline
\end{tabular}\\[2pt]
\end{small}
\end{center}
\end{table*}

\begin{figure}[tbh]
\vspace*{5.5cm}
\includegraphics{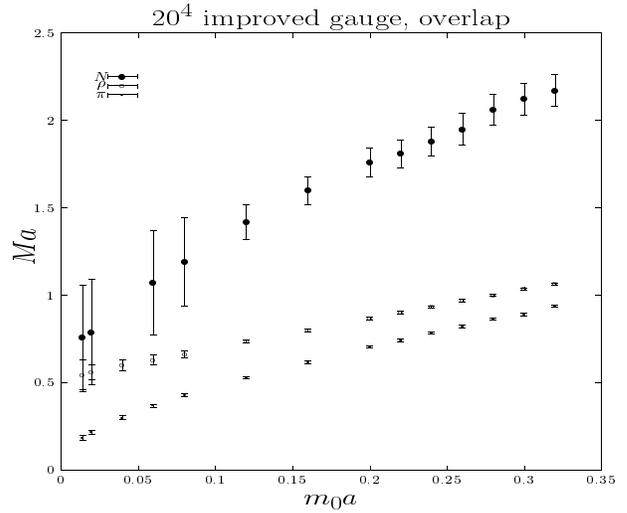} \caption{Nucleon, vector and pseudoscalar meson masses.}
\label{hmass} \end{figure}

    Next we plot the nucleon, the vector and pseudoscalar masses as a
function of $m_0a$ in Figure~\ref{hmass}. A simple linear fit with 10 smallest quark
masses gives  $m_V a = 0.534(12)$ at the chiral limit with  $\chi^2/DF
 = 0.05$. Using the $r_0$ to set the scale, this corresponds
to 712(16) MeV. Since the error in $m_N$ is still large, we can not
draw any conclusion on the $m_N/m_{\rho}$ ratio yet.

\begin{table*}[ht]
\caption{The renormalized quark mass $\frac{Z_P(\mu)}{Z_A(\mu)}m_q^{\overline{MS}} 
(\mu)a$ fitted in the form $c + A m_0a + B m_0^2a^2$.} \label{table:2} 
\newcommand{\m}{\hphantom{$-$}}
\newcommand{\cc}[1]{\multicolumn{1}{c}{#1}}
\renewcommand{\tabcolsep}{2pc} 
\renewcommand{\arraystretch}{1.2} 
\begin{center}
\begin{small}
\begin{tabular}{@{}llllll}
\hline
\#  & c & A  & B & $\chi^2$/DF  & $Z_A(\mu,0)$\\ 
\hline   
   4 & -0.00041(34) & 0.939(24) & 1.66(34) & 0.003 &  1.065(27) \\
   5 & -0.00049(24) & 0.945(14) & 1.56(16) & 0.05  &  1.058(16) \\
   6 &  -0.00063(16)& 0.956(8)  & 1.43(7)  & 0.29  &  1.046(9)   \\
\hline
\end{tabular}\\[2pt]
\end{small}
\end{center}
\end{table*}

\begin{figure}[tbh]
\vspace*{5.0cm}
\includegraphics{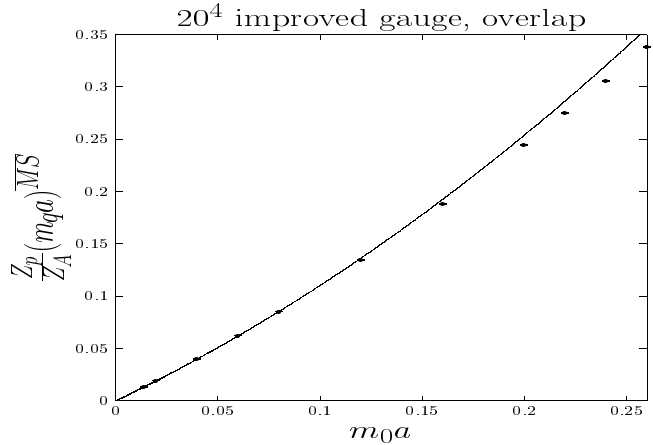}
\caption{The quark mass in the $\overline{MS}$ scheme. The solid line is a 
linear plus quadratic fit with the 4 lightest mass points.}
\label{qmass}
\end{figure}

   We have computed the quark mass through the chiral Ward
identity $Z_A\partial_{\mu} A_{\mu} = 2 Z_S^{-1}m_0 Z_P P$ where 
$A_{\mu}= \bar{\psi}i\gamma_{\mu}\gamma_5(\tau/2)\psi$ and
$P = \bar{\psi}i\gamma_5(\tau/2)\psi$. We plot
$\frac{Z_P(\mu)}{Z_A(\mu)}m_q^{\overline{MS}} (\mu) = Z_A^{-1}(\mu) m_0$
in Figure~\ref{qmass} and fit them in the form $c + A m_0a + B m_0^2a^2$. The
fits with a few smallest quark masses are given
in Table~\ref{table:2}. We see that the intercept $c$ is consistent with zero for the
case with 4 smallest masses. From this we can deduce the
non-perturbative renormalization constant $Z_A(\mu, m_0)$ as a function of
$m_0a$, i.e. $Z_A^{-1}(\mu, m_0) = A + B m_0 a$.  We see from Table~\ref{table:2} that
$Z_A(\mu, m_0 =0)$ is fairly far from the tree-level value of 1.368.

    Finally we shall explore the behavior of the pseudoscalar meson mass on
a small volume. Starting from the generalized Gell-Mann-Oakes-Renners relation
\begin{equation}   \label{gor}
m_0a\int d^4x \langle \pi(x) \pi(0)\rangle = 2 \langle\bar{\psi}\psi\rangle,
\end{equation}
and assuming pion dominance, one obtains the usual Gell-Mann-Oakes-Renners 
relation
\begin{equation}   \label{gor_usual}
m_{\pi}^2 = \frac{2 m_0}{f_{\pi}^2} \langle\bar{\psi}\psi\rangle.
\end{equation}
In the quenched approximation, the chiral condensate has the following
behavior for small $m_0$
\begin{equation}
\langle\bar{\psi}\psi\rangle = \frac{\langle |Q|\rangle}{m_0 V} + a + b\, m_0,
\end{equation}

\begin{figure}[th]
\vspace*{12.0cm}
\includegraphics{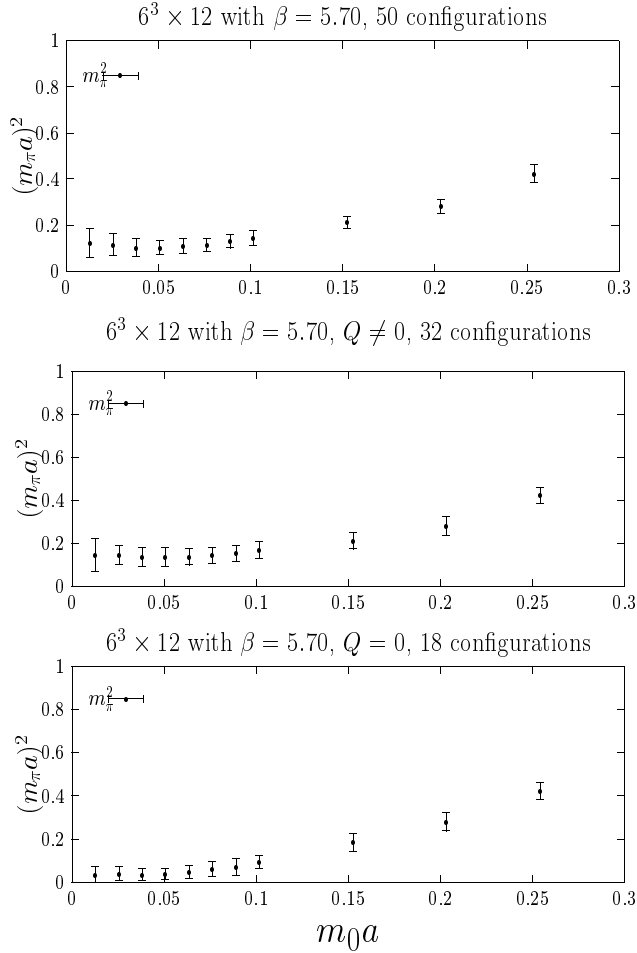} 
\caption{$m_{PS}^2a^2$ as a function of $m_0a$.}
\label{pi2Q3} 
\end{figure}
\noindent
where the first term due to the fermion zero modes associated with the topological 
charge $Q$ is the quenched artifact which is prominent as $m_0 \rightarrow 0$
and/or at small volume V. This term is observed in
$\langle\bar{\psi}\psi\rangle$ through a direct calculation with the overlap
fermion~\cite{ldl00}. Thus, barring additional complication due to the
quenched chiral log, the pion mass is expected to have the following dependence
on the quark mass $m_0$
\begin{equation}  \label{piQ}
m_{\pi}^2  =\frac{2}{f_{\pi}^2}\{\frac{\langle |Q|\rangle}{V} + a m_0 + b\,
m_0^2\}.
\end{equation}

We display in Figure~\ref{pi2Q3} $m_{\pi}^2a^2$ as a function of $m_0a$ for the $6^3 \times 
12$ lattice with 50 Wilson gauge configurations at $\beta = 5.7$. The top one
shows that $m_{\pi}^2$ approaches a constant in the chiral limit as shown
in Eq. (\ref{piQ}). To further verify that the constant is indeed due to the
nonvanishing topological charge in the configuration, we separate the 50 gauge
configurations into 32 $Q \neq 0$ ones (the middle figure) and 18 with $Q = 0$
(the bottom figure). We see that $m_{\pi}^2 a^2$ for the $Q \neq 0$ case 
approaches the same constant as $m_0a \rightarrow 0$, while it is much smaller
for the  $Q = 0$ case which has a tendency to approach zero at the chiral limit.
From the curvature of the dependence on $m_0a$, one speculates that $m_0a$ is
below the Thouless energy in such a small volume so that the $a m_0$ term in
Eq. (\ref{piQ}) vanishes. As a result, $m_{\pi}^2$ is proportional to $m_0^2$
and consequently the pion mass approaches the chiral limit linearly with $m_0$.
However, one needs more statistics to confirm this scenario.

To conclude, we demonstrate that it is feasible to implement the overlap
fermions on large physical volumes.
We do not see the quenched chiral log in the range $m_{\pi}/m_{\rho}
\sim 0.35 - 0.80$. 

  This work is partially supported by
U.S. DOE Grants DE-FG05-84ER40154 and DE-FG02-95ER40907.

\end{document}